\documentclass[letterpaper,journal]{IEEEtran}

\pdfoutput=1
\usepackage{cite}
\usepackage{amsmath,amsfonts}
\usepackage{algorithmic}
\usepackage{algorithm}
\usepackage{array}
\usepackage[caption=false,font=footnotesize,labelfont=rm,textfont=rm,subrefformat=parens]{subfig}
\usepackage{textcomp}
\usepackage{stfloats}
\usepackage{url}
\usepackage{verbatim}
\usepackage{graphicx}
\usepackage{xcolor}
\usepackage{multirow}
\usepackage{booktabs}
\usepackage[colorlinks, linkcolor=blue, anchorcolor=blue, citecolor=blue]{hyperref}

\makeatletter
\renewcommand{\paragraph}{\@startsection{paragraph}{4}{\parindent}%
	{0ex plus 0.1ex minus 0.1ex}{-0.0em}{\normalfont\normalsize\itshape}}
\makeatother
\begin{document}
	\title{Hierarchical Wireless Foundation Model for Multi-Task Optimization}
	
	\author{Yangjing Wang, Ouya Wang, Shenglong Zhou, and Geoffrey Ye Li
	\thanks{Yangjing Wang, Ouya Wang, and Geoffrey Ye Li are with the Department of Electrical and Electronic Engineering, Faculty of Engineering, Imperial College London, SW7 2BX London, U.K. (e-mail: {\{yangjing.wang25, ouya.wang20, geoffrey.li\}@imperial.ac.uk}). }
	\thanks{Shenglong Zhou is with the School of Mathematics and Statistics, Beijing Jiaotong University, China (e-mail: shlzhou@bjtu.edu.cn). \textit{(Corresponding author: Ouya Wang.)}}
	}
	
	\maketitle
	\begin{abstract}
		The increasing complexity of next-generation wireless networks has driven the integration of artificial intelligence (AI) into wireless communications. However, most existing studies focus on developing task-specific deep learning techniques for single scenarios, which limits their ability to generalize across diverse tasks, channel conditions, and system configurations. To address this generalization bottleneck, we propose a hierarchical wireless foundation model (WFM) for multi-task optimization. The proposed WFM couples an upstream foundation channel encoder (FCE) with a downstream foundation optimization decoder (FOD) via geometry-aware cross-attention. Specifically, the FCE extracts task-agnostic channel representations via self-supervised masked reconstruction while the FOD generates multi-task optimization decisions through differentiable output heads. Moreover, a hybrid supervised-to-unsupervised training strategy is employed to overcome the performance ceiling of purely supervised learning, and the modular architecture of the WFM enables efficient adaptation to unseen communication tasks with minimal parameter overhead. Simulation results show that the proposed WFM learns high-fidelity channel representations and achieves competitive multi-task optimization performance while substantially reducing optimization inference latency relative to numerical baselines. Furthermore, it exhibits robust generalization to unseen propagation environments, varying constraint parameters, and heterogeneous system configurations.
	\end{abstract}


	\section{Introduction}	
	The upcoming sixth-generation (6G) wireless networks aim to provide advanced services, such as ubiquitous connectivity and ultra-reliable low-latency communications. To support these services, 6G networks need to integrate massive antenna arrays and exploit extremely wide bandwidths at high frequencies, such as millimeter wave (mmWave) and terahertz (THz). However, these advancements create significant challenges for conventional models and algorithms. Massive antenna arrays introduce prohibitive computational complexity, rendering traditional approaches highly inefficient; operating at such high frequencies breaks the narrowband and far-field propagation assumptions these conventional approaches depend on, causing them to fail entirely \cite{saadVision6GWireless2020, daiDeepLearningWireless2020}.
	
	To overcome these challenges, recent studies have explored a variety of deep learning (DL) approaches. DL models drastically reduce computational overhead by mapping complex network observations directly to decision variables. Moreover, by learning entirely from data, DL models do not rely on any assumptions required by conventional approaches \cite{yePowerDeepLearning2018, qinDeepLearningPhysical2019, wang2022learn}. However, these models are often designed for task-specific scenarios, as different downstream applications necessitate distinct modeling strategies. This fragmentation increases the complexity of deployment, limits scalability and generality in practical settings, and is highly inefficient and time-consuming. While knowledge transfer and fine-tuning allow a single model to adapt to multiple tasks, they only perform well when those tasks are highly similar \cite{vanhuynhTransferLearningSignal2022, wangFrameworksFewshotLearning2025, wang2025fast}. These limitations highlight the need for a general and scalable wireless learning framework that can support diverse tasks and network settings, rather than rely on fragmented task-specific models.

	Here, \emph{scalability} means that the system maintains strong performance and manageable computational costs as downstream tasks (e.g., channel estimation, beamforming) change, or as data dimensions vary across system configurations (e.g., users, antennas, and subcarriers). Without sufficient scalability, handling heterogeneous multi-task settings or massive data dimensions would be computationally intractable, making real-world 6G deployment infeasible. Moreover, \emph{generality} means that the system maintains robust and accurate performance across diverse downstream tasks, channel distributions, and system constraints. Strong generality enables a ``train once, deploy everywhere'' paradigm: a single model architecture can adapt to multiple tasks, channels, or constraints with minimal fine-tuning, or even be applied directly with no retraining.

	Foundation models have emerged as a highly effective paradigm for achieving both scalability and generality. These large-scale, pre-trained backbones can be efficiently adapted to diverse tasks. For example, large language models (LLMs) serve as foundation models for diverse language processing tasks. Their adaptability comes from a shared and generalized backbone that extracts universal features, combined with lightweight, task-specific modules for downstream applications \cite{liangLargeLanguageModels2026, guoLargeAIModels2026}. Inspired by this paradigm, the wireless foundation model (WFM) has been proposed to handle multiple physical-layer wireless tasks within a single framework. 
	
	Recent studies on WFMs can be divided into two categories, as shown in Table~\ref{tab:foundation_models}. The first category is Wireless LLM, which uses pre-trained LLM architectures as the feature extraction backbone. The key technique enabling LLMs to perform well in wireless tasks is mapping physical wireless data (e.g., coordinates and azimuths for beam selection) directly into the discrete LLM token space \cite{shengBeamPredictionBased2025}. This mapping translates diverse wireless data into a common language that the LLM natively understands, allowing the same LLM backbone to extract different task features simply by changing the input prompts. However, mapping continuous wireless signals into discrete textual tokens creates an inherent modality gap that degrades performance. Moreover, the massive inference overhead of LLMs remains prohibitive for real-time deployment. 

	\begin{table*}[t]
		\centering
		\caption{Comparison of Existing Foundation Models and the Proposed Framework}
		\label{tab:foundation_models}
		\renewcommand{\arraystretch}{1.23}
		\resizebox{\textwidth}{!}{%
		\begin{tabular}{@{}llcllcc@{}}
		\toprule
		\textbf{Models} & \textbf{Refs.} & \textbf{Class} & \textbf{Target Tasks} & \textbf{Key Architecture} & \textbf{Contributions} & \textbf{Limitations} \\ \midrule

		\multicolumn{7}{@{}l}{\textit{Category I: Text Tokens - Wireless Data Alignment}} \\ \midrule
		Channel prediction LLM & \cite{liuLLM4CPAdaptingLarge2024} & Wireless LLM & Channel prediction & Pre-trained LLM + CSI embedding & a, c & 1, 2, 4 \\
		Beam prediction LLM & \cite{shengBeamPredictionBased2025} & Wireless LLM & Beam prediction & Pre-trained LLM + Tokenizer & a, c & 1, 2, 4 \\
		Multi-task framework & \cite{zhengLargeLanguageModel2026} & Wireless LLM & Heterogeneous tasks & Prompt-as-prefix conditioning & a, b & 1, 2, 3 \\
		MUSE-FM & \cite{zhengMUSEFMMultitaskEnvironmentaware2025} & Wireless LLM & Cross-scenario tasks & Multimodal contextual inputs & b, c, d & 1, 2, 3  \\ \midrule

		\multicolumn{7}{@{}l}{\textit{Category II: Wireless Data Pre-training}} \\ \midrule
		WiFo & \cite{liuWiFoWirelessFoundation2025} & WFM & Channel prediction & Masked autoencoder (MAE) & a, c & 4, 5 \\
		WiFo-CF & \cite{liuWiFoCFWirelessFoundation2025} & WFM & CSI feedback & Mixture-of-experts (MoE) & a, c & 4, 5 \\
		Multi-task WFM & \cite{shengWirelessFoundationModel2025} & WFM & Channel, angle, traffic pred. & Univariate decomposition & b, c, d & 4, 5 \\
		Task-agnostic FCM & \cite{alikhaniLargeWirelessModel2025} & FCM & Representation learning & Masked channel modeling & d, e & 5, 6, 7 \\
		WirelessGPT & \cite{yangWirelessGPTGenerativePreTrained2025} & FCM & Representation learning & Tri-domain attention fusion & d, e & 3, 5, 6, 7 \\
		App-oriented FCM & \cite{jingSignalCompressionWireless2026} & FCM & Regression \& classification & Masked reconstruction & d, e &5, 6, 7 \\ 

		\textbf{Hierarchical WFM} & \textbf{Ours} & \textbf{Hierarchical WFM} & \textbf{Multi-task optimization} & \textbf{Asymmetric FCE + Prompt FOD} & \textbf{b, c, d, e, f, g} & \textbf{-} \\ \bottomrule

		\multicolumn{7}{@{}p{\textwidth}@{}}{
		\vspace{0.1mm}
		\footnotesize \textbf{Contributions:} \textbf{(a)} Single-task performance enhancement; \textbf{(b)} Multi-task learning; \textbf{(c)} New scenario/configuration generalizability; \textbf{(d)} New task adaptation; \textbf{(e)} Universal channel representation learning; \textbf{(f)} Accelerated inference; \textbf{(g)} Multi-task optimization capability.} \\

		\multicolumn{7}{@{}p{\textwidth}@{}}{
		\vspace{0.1mm}
		\footnotesize \textbf{Limitations:} \textbf{(1)} Modality gap between textual tokens and wireless signals; \textbf{(2)} Prohibitive inference overhead for real-time deployment; \textbf{(3)} Suboptimal performance due to causal self-attention; \textbf{(4)} Restriction to predefined task clusters; \textbf{(5)} Failure to fully exploit the spatial-frequency redundancy of CSI; \textbf{(6)} Neglect of phase fidelity; \textbf{(7)} Lack of explicit mechanisms to model the multi-user dimension.}
		\end{tabular}
		}
	\end{table*}

	To overcome these limitations, the second category of research focuses on native WFMs. These native WFMs have lightweight architectures and are pre-trained from scratch using wireless data, thereby reducing inference overhead and eliminating the modality gap. Early WFMs focused on single-task generalization across varying system configurations, utilizing a masked autoencoder (MAE) framework for channel prediction \cite{liuWiFoWirelessFoundation2025} and a mixture-of-experts (MoE) architecture for channel state information (CSI) feedback \cite{liuWiFoCFWirelessFoundation2025}. Subsequent work has extended WFMs to multi-task paradigms, which unify heterogeneous prediction tasks and demonstrate zero-shot generalization to new tasks \cite{shengWirelessFoundationModel2025}. However, their zero-shot generalization degrades on tasks dissimilar to those seen during pre-training. This degradation occurs because these WFMs tightly couple general channel features with specific task objectives. To address this challenge, foundation channel models (FCMs) have been developed to decouple universal channel feature extraction from downstream task objectives. This decoupling allows them to extract universal, task-agnostic features, achieving significantly broader cross-domain generalizability for diverse multi-task applications \cite{alikhaniLargeWirelessModel2025}. 

	Native WFMs are well-suited for physical-layer multi-task optimization. Conventionally, these optimization tasks are tackled by isolated algorithms or task-specific DL models. Because these tasks share the same propagation environment, processing them independently forces the system to repeatedly extract identical environmental features. This redundant computation creates a severe bottleneck, causing massive delays that violate strict 6G real-time latency requirements. To overcome this, we must exploit the intrinsic correlation among these tasks. Despite their distinct mathematical objectives, these tasks share a dependence on the same wireless environment, naturally motivating a unified optimization foundation model. By extracting a universal representation just once and reusing it across multiple optimization tasks within a single framework, this approach eliminates redundant computations, substantially reduces processing latency, and enables practical real-time deployment.

	However, existing WFMs face critical limitations when applied to physical-layer multi-task optimization because they
	\begin{itemize}		
		\item target narrow sets of related tasks under fixed system configurations, thereby limiting their adaptability and generalizability across varying constraints and heterogeneous system configurations;

		\item rely on causal self-attention, which restricts attention to only past tokens rather than the full channel observation context, thereby rendering these models suboptimal for instantaneous optimization decisions;
		
		\item process CSI through encoder-only architectures, which leaves the high spatial-frequency redundancy underexploited, thereby significantly increasing inference latency and preventing real-time decision-making;
		
		\item represent CSI solely over the spatial-frequency domain without an explicit user dimension, which limits their ability to capture inter-user coupling and generalize across varying user counts in multi-user optimization;
		
		\item focus on amplitude reconstruction while neglecting phase fidelity in their pre-training objectives, which discards essential directional signal information, thereby degrading performance on phase-sensitive tasks.
	\end{itemize}
	Therefore, we develop a hierarchical WFM to address these limitations systematically.

	\section{Contributions}
	In this work, we propose a hierarchical WFM designed to meet the scalability and generality requirements of next-generation networks. Scalability is achieved through a modular architecture that supports adaptation to new tasks and system configurations with minimal overhead, while generality is enabled by a pre-trained backbone that performs robustly across diverse tasks and unseen conditions. This unified design offers both low-latency inference and strong adaptability, providing a practical foundation for 6G networks. The key contributions are fourfold.

	\begin{itemize}
		\item \textbf{Achieving scalability and generality via a hierarchical architecture:} 
		We propose a hierarchical architecture that decouples channel feature extraction from task-specific optimization. Specifically, an upstream foundation channel encoder (FCE) extracts universal channel features, while a downstream foundation optimization decoder (FOD) maps these features to task-specific decisions. Therefore, this framework directly achieves scalability by maintaining low computation and inference overhead as the number of downstream tasks grows, and generality by supporting tasks with fundamentally different objectives and constraints.

		\item \textbf{Learning universal channel features with efficiency and phase fidelity:} 
		The upstream FCE adopts an MAE framework with a high masking ratio to fully exploit the spatial-frequency redundancy of CSI. To accommodate a variable number of users, a per-user patching scheme decouples the permutation-invariant user dimension from the spatial-frequency grid. Furthermore, a phase-sensitive pre-training scheme jointly preserves amplitude and phase fidelity. Therefore, the FCE efficiently learns universal channel representations that generalize across varying user counts and retain the directional information essential for phase-sensitive downstream tasks.

		\item \textbf{Enabling multi-task optimization through prompt-conditioned cross-attention:}
		The downstream FOD encodes task identities and user-specific constraints into a composite prompt, which queries universal channel features via geometry-aware cross-attention. Differentiable output heads then map the decoded representations to physically feasible solutions. Therefore, the FOD supports heterogeneous tasks within a single decoder, enables zero-shot generalization across varying system configurations, and produces instantaneous optimization decisions conditioned on the full channel observation.

		\item \textbf{Validating scalability and generality via extensive experiments:}
		Experiments across four representative tasks validate the scalability and generality of the proposed WFM. For scalability, the WFM adapts to new tasks with minimal parameter overhead and achieves substantially lower inference latency than numerical baselines. For generality, it delivers competitive multi-task performance and robust zero-shot generalization to unseen propagation environments, varying constraint parameters, and heterogeneous system configurations. Ablation studies further confirm the effectiveness of its key design components.
	\end{itemize}

	\section{System Model and Problem Formulation}
	We consider a multi-user multiple-input single-output orthogonal frequency-division multiplexing (MU-MISO-OFDM) system operating in time-division duplex (TDD) mode, where a base station (BS) equipped with $M$ antennas serves $K$ single-antenna users over $N$ orthogonal subcarriers. Let $\mathbf{h}_{k,n} \in \mathbb{C}^{M}$ denote the downlink channel vector from the BS to user $k$ on subcarrier $n$, and define $\mathbf{w}_{k,n} \in \mathbb{C}^{M}$ as the corresponding precoding vector. The aggregate CSI tensor over all users and subcarriers is denoted as $\mathbf{H}\in\mathbb{C}^{K\times N\times M}$, and the index set of users is expressed as $\mathcal{K}=\{1, \ldots, K\}$.

	Based on the system model, the received signal at user $k$ on subcarrier $n$ is given by
	\begin{equation}
	y_{k,n} = \mathbf{h}_{k,n}^H \mathbf{w}_{k,n} x_{k,n} + \mathbf{h}_{k,n}^H \sum_{j \neq k} \mathbf{w}_{j,n} x_{j,n} + z_{k,n},
	\end{equation}
	where $x_{k,n}$ denotes the transmitted data symbol for user $k$ on subcarrier $n$, satisfying $\mathbb{E}[|x_{k,n}|^2]=1$, and $z_{k,n}\sim\mathcal{CN}(0,\sigma^2)$ is the corresponding additive white Gaussian noise (AWGN). The noise variance is assumed identical across users and subcarriers. Thus, the signal-to-interference-plus-noise ratio (SINR) for user $k$ on subcarrier $n$ is obtained as
	\begin{equation}
	\gamma_{k,n} = \frac{|\mathbf{h}_{k,n}^H \mathbf{w}_{k,n}|^2}{\sum_{j \neq k} |\mathbf{h}_{k,n}^H \mathbf{w}_{j,n}|^2 + \sigma^2},
	\end{equation}
	and the corresponding achievable rate of user $k$ is written as
	\begin{equation}
	R_k = \sum_{n=1}^{N} \log_2(1 + \gamma_{k,n}).
	\end{equation}

	\subsection{Beamforming}
	Under the assumption of perfect CSI at the BS, the beamforming task aims to maximize the system sum rate by jointly optimizing the precoding vectors across all users and subcarriers. The optimization problem is formulated as
	\begin{subequations}
		\begin{align}
		& \max_{\mathbf{W}} \quad \sum_{k=1}^{K} R_k, \\
		& \ \text{s.t.} \hspace{0.5cm} \sum_{k=1}^{K} \sum_{n=1}^{N} \|\mathbf{w}_{k,n}\|^2 \leq P_{\max},
		\end{align}
	\end{subequations}
	where $\mathbf{W} \in \mathbb{C}^{K \times N \times M}$ represents the aggregate precoding tensor, and $P_{\max}$ is the maximum transmission power at the BS. Due to the non-concavity of the objective, conventional iterative algorithms typically converge to locally optimal solutions and incur computational costs that scale unfavorably with $K$ and $N$, motivating learning-based approaches.

	\subsection{User Scheduling}
	When the number of users exceeds that of BS antennas, serving all users simultaneously induces severe multi-user interference, motivating the selection of an active subset for transmission. Let $s_k \in \{0, 1\}$ denote the scheduling indicator of user $k$, with global scheduling vector $\mathbf{s} = [s_1, \ldots, s_K]^T$. For the scheduled users, we adopt regularized zero-forcing (RZF) precoding with equal power allocation as the fixed scheme. The SINR for user $k$ on subcarrier $n$ is given by
	\begin{equation}
	\gamma_{k,n}(\mathbf{s})=\frac{s_k |\mathbf{h}_{k,n}^H\mathbf{w}_{k,n}|^2}{\sum_{j\neq k}s_j|\mathbf{h}_{k,n}^H\mathbf{w}_{j,n}|^2+\sigma^2},
	\end{equation}
	and the corresponding achievable rate of user $k$ is written as
	\begin{equation}
	R_k(\mathbf{s}) = \sum_{n=1}^{N}\log_2\left(1+\gamma_{k,n}(\mathbf{s})\right).
	\end{equation} 
	Thus, the user scheduling problem is formulated as
	\begin{subequations}
		\begin{align}
		& \max_{\mathbf{s}} \quad \sum_{k=1}^{K} R_k(\mathbf{s}), \\
		& \ \text{s.t.} \hspace{0.5cm} s_k R_k(\mathbf{s}) \geq s_k R_{\min}, \quad \forall k \in \mathcal{K}, \\
		& \hspace{1.05cm} s_k \in \{0, 1\}, \quad \forall k \in \mathcal{K},
		\end{align}
	\end{subequations}
	where $R_{\min}$ denotes the minimum rate requirement for each scheduled user to satisfy its quality-of-service (QoS) constraint. The binary scheduling variables render the problem combinatorial with an exhaustive search complexity of $\mathcal{O}(2^K)$.

	\subsection{Channel Estimation}
	By exploiting TDD channel reciprocity, the BS acquires the downlink CSI via uplink pilot transmissions. Let $\mathcal{P} \subseteq \{1, \ldots, N\}$ denote the index set of pilot subcarriers with cardinality $N_p = |\mathcal{P}|$, defining the pilot ratio as $r_p = N_p/N$. With orthogonal pilot sequences across users and unit pilot transmission power, the interference-free observation for user $k$ on subcarrier $n$ can be obtained as
	\begin{equation}
	\mathbf{y}_{k,n} = \mathbf{h}_{k,n} + \mathbf{n}_{k,n},
	\end{equation}
	where $\mathbf{n}_{k,n} \sim \mathcal{CN}(\mathbf{0}, \sigma_p^2 \mathbf{I}_M)$ denotes the effective AWGN. The corresponding least-squares (LS) estimate is given by
	\begin{equation}
	\mathbf{h}_{k,n}^{\text{LS}} = \begin{cases} \mathbf{y}_{k,n}, & n \in \mathcal{P}, \\ \mathbf{0}, & n \notin \mathcal{P}. \end{cases}
	\end{equation}
	Stacking these estimates over all users and subcarriers yields LS estimate tensor $\mathbf{H}_{\text{LS}} \in \mathbb{C}^{K \times N \times M}$, which is sparse along the frequency dimension. The channel estimation task aims to recover global CSI tensor $\mathbf{H}$ from $\mathbf{H}_{\text{LS}}$, and the accuracy of recovered estimate $\hat{\mathbf{H}}$ is evaluated by the normalized mean-squared error (NMSE) with respect to $\mathbf{H}$. This task can be viewed as a joint denoising and interpolation problem over the spatial-frequency domain.

	\subsection{Beam Selection}
	In analog and hybrid beamforming architectures, the analog beamforming vectors are typically constrained to frequency-flat beams selected from a finite codebook since the analog phase-shifting network cannot realize arbitrary frequency-selective precoding. Assuming a uniform linear array (ULA) at the BS, we adopt a DFT codebook $\mathcal{C} = \{\mathbf{c}_0, \ldots, \mathbf{c}_{|\mathcal{C}|-1}\}$ with oversampling factor $o_s$ and cardinality $|\mathcal{C}| = o_s M$, where
	\begin{equation}
	\mathbf{c}_i = \frac{1}{\sqrt{M}} \left[1, e^{j2\pi i / |\mathcal{C}|}, \ldots, e^{j2\pi i(M-1) / |\mathcal{C}|}\right]^T.
	\end{equation}
	Let $i_k$ denote the codeword index assigned to user $k$, and $\mathbf{i}=[i_1,\ldots,i_K]^T$. Under equal power allocation, the SINR for user $k$ on subcarrier $n$ is obtained as
	\begin{equation}
	\gamma_{k,n}(\mathbf{i}) = \frac{|\mathbf{h}_{k,n}^H \mathbf{c}_{i_k}|^2}{\sum_{j \neq k} |\mathbf{h}_{k,n}^H \mathbf{c}_{i_j}|^2 + KN\sigma^2 / P_{\max}},
	\end{equation}
	and the corresponding achievable rate of user $k$ is written as
	\begin{equation}
	R_k(\mathbf{i}) = \sum_{n=1}^{N} \log_2 (1 + \gamma_{k,n}(\mathbf{i})).
	\end{equation}
	Thus, the beam selection problem is formulated as
	\begin{subequations}
	\begin{align}
	& \max_{\mathbf{i}} \quad \sum_{k=1}^{K} R_k(\mathbf{i}), \\
	& \ \text{s.t.} \hspace{0.5cm} i_k \in \{0, \ldots, |\mathcal{C}|-1\}, \quad \forall k \in \mathcal{K}.
	\end{align}
	\end{subequations}
	This discrete combinatorial assignment has a search space of $|\mathcal{C}|^K$, rendering exhaustive search computationally prohibitive as $|\mathcal{C}|$ and $K$ grow.

	Each of the above four tasks has been extensively studied using conventional algorithms, including weighted minimum mean-square error (WMMSE) precoding for beamforming \cite{shiIterativelyWeightedMMSE2011}, zero-forcing-based greedy selection for user scheduling \cite{yooOptimalityMultiantennaBroadcast2006}, LS and linear MMSE (LMMSE) estimators for channel estimation \cite{vandebeekChannelEstimationOFDM1995}, and hierarchical codebook search for beam selection \cite{alkhateebChannelEstimationHybrid2014}. More recently, task-specific DL models have been developed to learn direct mappings from channel observations to task-specific outputs \cite{yePowerDeepLearning2018, xiaDeepLearningFramework2020}. However, both lines of work address each task independently, leaving the shared channel structure underlying these tasks unexploited.

	\section{Hierarchical Wireless Foundation Model}
	\begin{figure*}[htbp]
		\centering
		\includegraphics[width=0.6\textwidth]{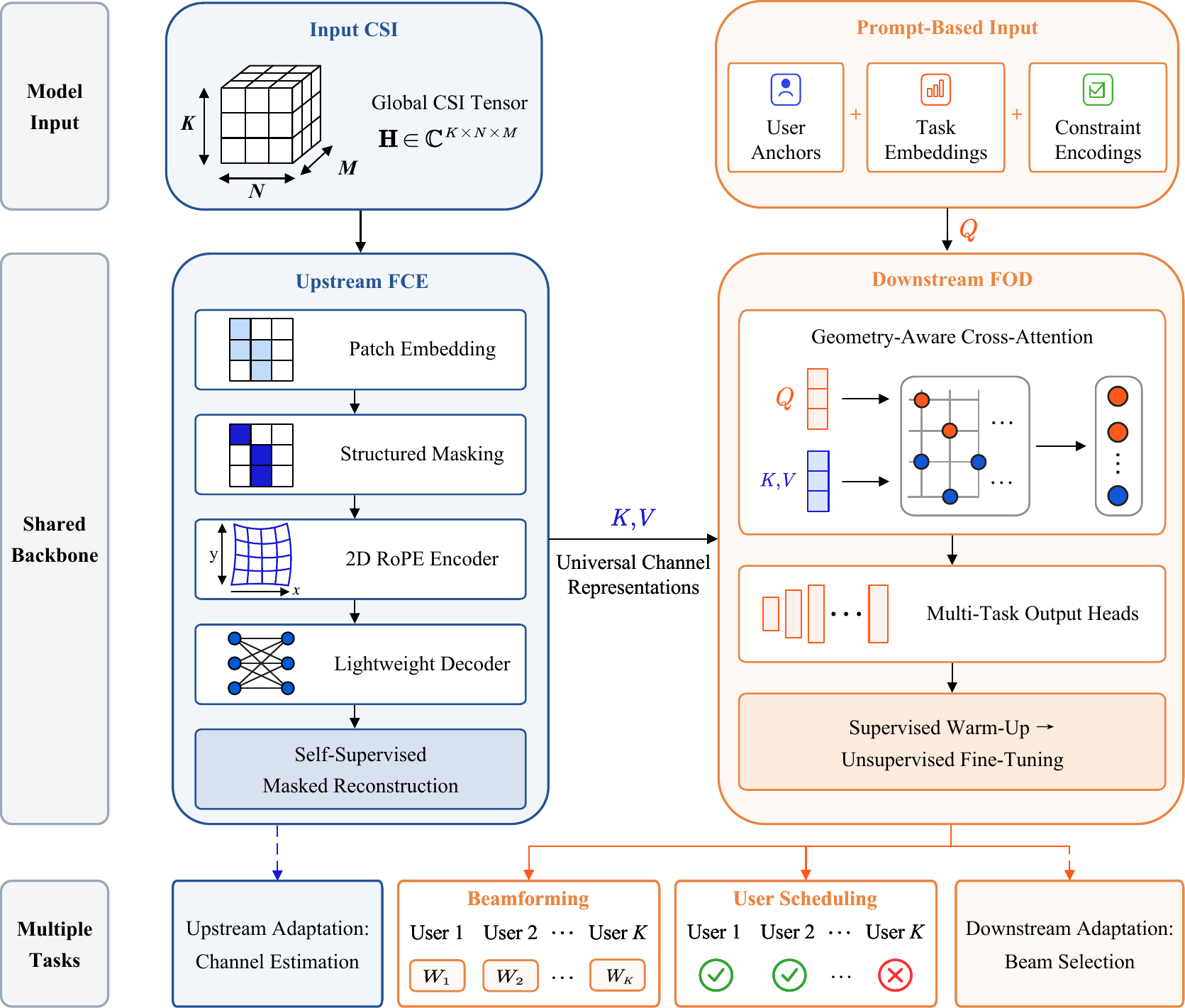}
		\caption{Overall workflow of the proposed hierarchical WFM.}
		\label{WFM}
	\end{figure*}

	Although the four tasks formulated in the previous section involve heterogeneous objectives and decision variables, they all rely on the same underlying CSI tensor $\mathbf{H}$. To exploit this shared dependence while accommodating task heterogeneity, the proposed WFM adopts a hierarchical architecture that decouples task-agnostic channel representation learning from task-conditioned decision-making.

	As illustrated in Fig.~\ref{WFM}, the upstream FCE encodes CSI tensor $\mathbf{H}$ into universal channel representations that are independent of any downstream objective. The downstream FOD constructs task-specific prompts from user anchors, task embeddings, and constraint encodings. Through geometry-aware cross-attention, these prompts extract task-relevant information from universal channel representations and generate decisions for the target task. Since task conditioning is confined to the FOD, each channel realization is encoded only once and reused across multiple downstream tasks, avoiding repeated feature extraction and reducing multi-task inference overhead.

	The overall framework is trained and extended in three stages. First, the FCE is pre-trained via self-supervised masked reconstruction on a multi-scenario CSI corpus. Second, with the pre-trained FCE frozen, the FOD is trained for beamforming and user scheduling using supervised warm-up followed by unsupervised fine-tuning. Third, the pre-trained backbone is extended to channel estimation through upstream FCE adaptation and to beam selection through downstream FOD adaptation, with only lightweight task-specific modules updated. The following three sections detail these stages in turn.

	\section{Foundation Channel Encoder}
\begin{figure}[htbp]
		\centering
		\includegraphics[width=0.43\textwidth]{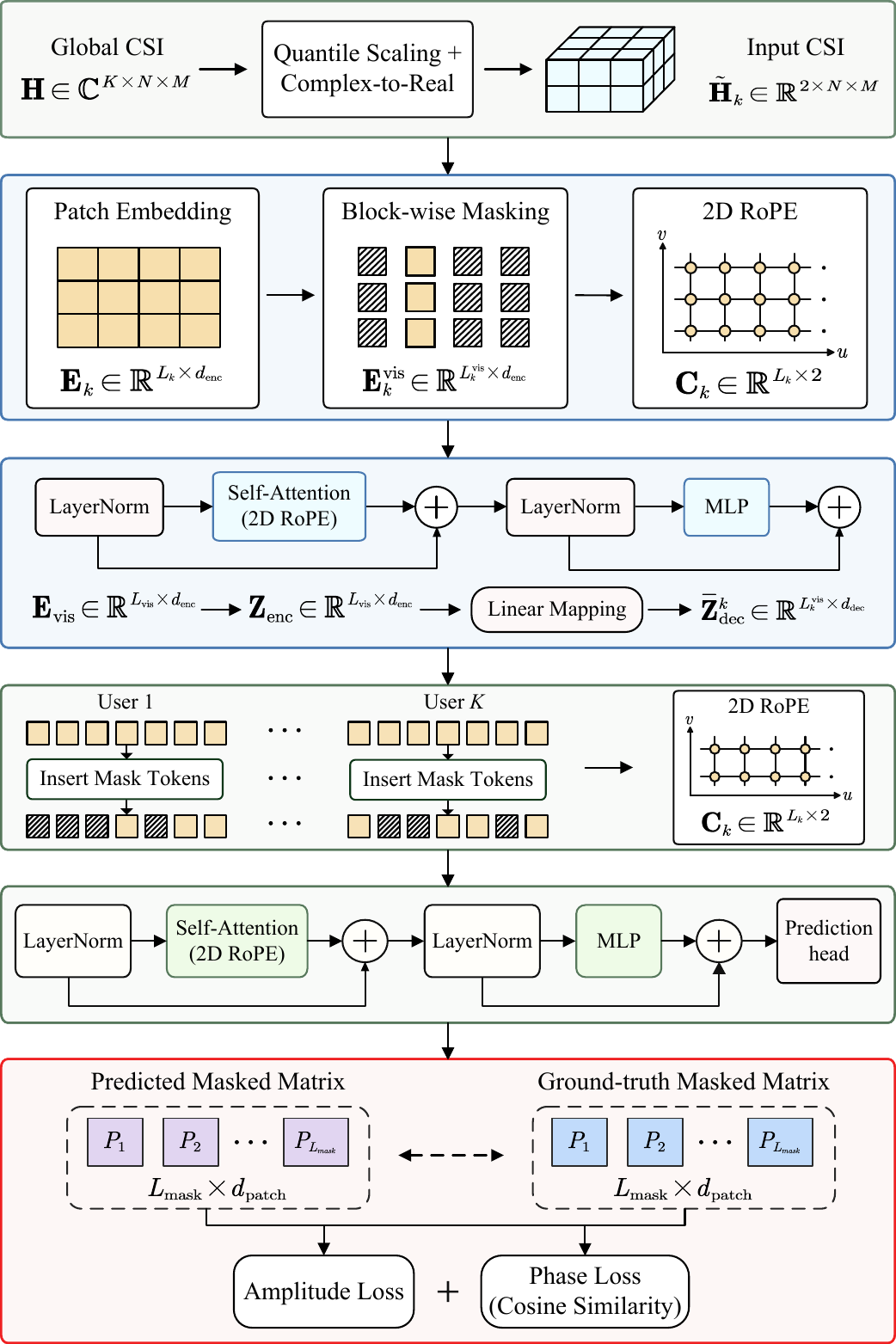}
		\caption{Network architecture of the proposed FCE.}
		\label{FCE}
	\end{figure}

	As the upstream representation learning module, the FCE extracts universal channel representations from global CSI tensor $\mathbf{H}$ via a self-supervised MAE framework \cite{heMaskedAutoencodersAre2022}, which exploits the high spatial-frequency redundancy of CSI. In this section, we first present the network architecture of the FCE, as shown in Fig.~\ref{FCE}, and then introduce a phase-sensitive training scheme that jointly optimizes amplitude and phase fidelity.

	\subsection{Input Preprocessing}
	Since channel amplitudes span several orders of magnitude across diverse propagation environments, raw CSI samples are rescaled to a numerically stable range to facilitate gradient-based optimization. We define a global scaling factor as the reciprocal of the $q$-quantile of the absolute amplitudes aggregated over the training corpus. As this quantile-based scaling is robust against outliers and preserves relative magnitude differences across users, antennas, and subcarriers, the large-scale fading information is retained for downstream tasks. The normalized complex-valued tensor is then split into real and imaginary components, yielding real-valued representation $\tilde{\mathbf{H}}$.

	\subsection{Per-User Patching and Embedding}
	In contrast to the structured correlations along the frequency and spatial dimensions, the user dimension exhibits inherent permutation invariance. Therefore, we apply patching to each user's channel independently. Following the patch-based tokenization paradigm \cite{dosovitskiyImageWorth16x162021}, each per-user tensor $\tilde{\mathbf{H}}_k \in \mathbb{R}^{2 \times N \times M}$ is partitioned into non-overlapping 2D patches of size $(n, m)$ along the frequency and spatial dimensions. Patching and linear projection are jointly implemented via a shared 2D convolution
	\begin{equation} \label{Conv}
	\mathbf{E}_k = \operatorname{Flatten}\Bigl(\operatorname{Conv2d}(\tilde{\mathbf{H}}_k)\Bigr),
	\end{equation}
	where $\operatorname{Conv2d}(\cdot)$ represents a 2D convolution operator with both kernel size and stride set to $(n,m)$, mapping two real-valued input channels to $d_{\text{enc}}$ output channels, and $\operatorname{Flatten}(\cdot)$ reshapes the convolutional output into $\mathbf{E}_k \in \mathbb{R}^{L_k \times d_{\text{enc}}}$ with $L_k = \frac{N}{n} \times \frac{M}{m}$. Since this shared convolutional operator is applied independently to each user, the per-user embeddings remain permutation-equivariant and naturally accommodate a variable number of users $K$ across samples.

	\subsection{Structured Masking Strategy}
	Due to the high spatial-frequency redundancy of CSI, random patch masking can allow masked patches to be trivially reconstructed via local interpolation, preventing the encoder from learning robust long-range representations. We therefore adopt a block-wise masking strategy \cite{baoBEiTBERTPreTraining2021} to mask contiguous blocks of patches, suppressing short-range information leakage and compelling the encoder to exploit the global spatial-frequency context.
	Moreover, applying a masking ratio jointly over all users' patches induces user-level variance in the masked proportion, leading to imbalanced supervision across users. To mitigate this effect, block-wise masking is performed independently for each user. Specifically, we iteratively sample non-overlapping blocks until the target masking ratio is attained, yielding a visible index set $\mathcal{I}_k^{\text{vis}}$ and a masked index set $\mathcal{I}_k^{\text{mask}}$ with $|\mathcal{I}_k^{\text{vis}}| + |\mathcal{I}_k^{\text{mask}}| = L_k$. Accordingly, user embedding $\mathbf{E}_k$ is partitioned as
	\begin{equation}
	\mathbf{E}_k^{\text{vis}} = \mathbf{E}_k[\mathcal{I}_k^{\text{vis}}], \quad \mathbf{E}_k^{\text{mask}} = \mathbf{E}_k[\mathcal{I}_k^{\text{mask}}],
	\end{equation}
	where $L_k^{\text{vis}} = |\mathcal{I}_k^{\text{vis}}|$, $L_k^{\text{mask}} = |\mathcal{I}_k^{\text{mask}}|$, $\mathbf{E}_k^{\text{vis}} \in \mathbb{R}^{L_k^{\text{vis}} \times d_{\text{enc}}}$ and $\mathbf{E}_k^{\text{mask}} \in \mathbb{R}^{L_k^{\text{mask}} \times d_{\text{enc}}}$ denote the visible and masked embeddings, respectively. The visible embeddings are forwarded to the encoder while the masked embeddings are discarded and the corresponding original CSI patches are retained as reconstruction targets during pre-training.

	\subsection{Shared Positional Embedding}
	Since self-attention without positional information is permutation-equivariant, positional encodings are required to inject the geometric structure of the spatial-frequency grid. To enable zero-shot generalization across heterogeneous configurations $(N, M)$, we employ a shared 2D rotary position embedding (RoPE) strategy that factorizes the frequency and spatial axes. We define a coordinate matrix $\mathbf{C} \in \mathbb{R}^{L_k \times 2}$ shared across users, where the $l$-th row $(u_l, v_l)$ denotes the frequency and spatial indices of the $l$-th patch. This shared coordinate matrix ensures that patches with identical physical roles receive identical positional encodings. 
	To encode the two axes independently, a feature vector $\mathbf{x}_l \in \mathbb{R}^{1 \times d_h}$ within an attention head of dimension $d_h$ is evenly split into $\mathbf{x}_l = [\mathbf{x}_{l,f}, \mathbf{x}_{l,s}]$, where $\mathbf{x}_{l,f}$ and $\mathbf{x}_{l,s}$ denote the frequency-selective and spatial-correlation features, respectively.
	Let $\mathbf{R}_{\Theta}(p)$ denote the standard 1D rotation matrix \cite{suRoFormerEnhancedTransformer2024}, parameterized by a position scalar $p$ and a rotary-frequency basis $\Theta$. The axis-factorized rotary embedding is then given by
	\begin{equation}
	f_{\text{RoPE}}(\mathbf{x}_l, u_l, v_l) = \operatorname{Concat}\!\left(\mathbf{x}_{l,f} \mathbf{R}_{\Theta}(u_l), \mathbf{x}_{l,s} \mathbf{R}_{\Theta}(v_l)\right).
	\end{equation}
	This factorization assigns independent embedding subspaces to the frequency and spatial axes, enabling the model to disentangle frequency selectivity from spatial correlations and adapt to arbitrary grid sizes $(N, M)$.

	\subsection{Encoder}
	To capture inter-user interference, the FCE encoder jointly processes the global visible embedding while its coordinate matrix provides patch coordinates for the subsequent 2D RoPE.
	The encoder comprises a stack of Transformer blocks \cite{vaswaniAttentionAllYou2017}, each integrating positional context into multi-head self-attention (MSA) via 2D RoPE. For the $i$-th head, the block input is linearly projected into the query, key, and value matrices $\mathbf{Q}_i$, $\mathbf{K}_i$, and $\mathbf{V}_i$, respectively. Based on the corresponding coordinate $(u_l, v_l)$, the query and key vectors of patch $l$ are rotated as
	\begin{equation}
	\tilde{\mathbf{q}}_{l,i} = f_{\text{RoPE}}(\mathbf{q}_{l,i}, u_l, v_l), \quad \tilde{\mathbf{k}}_{l,i} = f_{\text{RoPE}}(\mathbf{k}_{l,i}, u_l, v_l).
	\end{equation}
	Stacking the rotated vectors into $\tilde{\mathbf{Q}}_i$ and $\tilde{\mathbf{K}}_i$, the scaled dot-product attention for head $i$ is formulated as
	\begin{equation} \label{head}
	\operatorname{head}_i = \operatorname{softmax} \biggl(\frac{\tilde{\mathbf{Q}}_i \tilde{\mathbf{K}}_i^T}{\sqrt{d_h}} \biggr) \mathbf{V}_i.
	\end{equation}
	The outputs of all heads are concatenated and linearly projected to form the MSA output. After passing through all blocks, the encoder yields context-aware universal representation $\mathbf{Z}_{\text{enc}}$, transforming independent user embeddings into a system-level representation through self-attention.

	\subsection{Decoder}
	The decoder reconstructs the masked CSI patches from the encoded representations. Unlike the encoder that jointly processes visible patches across all users, the decoder operates on each user's representation independently. This asymmetric design forces the encoder to capture inter-user interference while confining the decoder to per-user reconstruction. To reinforce this asymmetry, the decoder is instantiated as a lightweight Transformer network with model dimension $d_{\text{dec}} \ll d_{\text{enc}}$.

	The global representation $\mathbf{Z}_{\text{enc}}$ is partitioned into per-user representations and linearly projected into the decoder feature space. A learnable mask token fills the masked positions, and the visible and masked tokens are restored to their original order according to $\mathcal{I}_k^{\text{vis}}$ and $\mathcal{I}_k^{\text{mask}}$. Then, the shared coordinate matrix is applied to 2D RoPE in the decoder MSA, and a linear prediction head maps the decoder output to the patch space. Masked-position predictions $\mathbf{Y}_k^{\text{mask}}$ are concatenated across users to form the reconstructed masked-patch matrix
	\begin{equation}
	\hat{\mathbf{H}}_m = \operatorname{Concat}\bigl(\mathbf{Y}_1^{\text{mask}}, \dots, \mathbf{Y}_K^{\text{mask}}\bigr).
	\end{equation}
	The corresponding ground-truth matrix $\mathbf{H}_m \in \mathbb{R}^{L_{\text{mask}} \times d_{\text{patch}}}$ is obtained by extracting and flattening the original normalized CSI patches at the same indices, where $L_{\text{mask}} = \sum_{k=1}^{K} L_k^{\text{mask}}$ and $d_{\text{patch}} = 2 \times n \times m$.

	\subsection{Pre-Training Strategy}
	The FCE is pre-trained via self-supervised reconstruction over a heterogeneous CSI corpus. Since system dimensions $(K, N, M)$ vary substantially across scenarios, we adopt a homogeneous batching strategy, where each mini-batch is drawn exclusively from a single $(K, N, M)$ subset to ensure dimensional consistency. These subsets are sampled with probabilities proportional to their corpus frequencies, thereby preserving the native configuration distribution and avoiding the over-representation of rare configurations.

	Regarding the optimization objective, the standard MSE is dominated by high-power channel entries and thus underweights the reconstruction errors of low-power entries, whose phase information remains critical to phase-sensitive downstream tasks, such as beamforming and channel estimation. To mitigate this bias, we complement the MSE with a cosine-similarity term, yielding the pre-training objective
	\begin{equation}
	\mathcal{L} = \frac{1}{N_{\text{mask}}} \|\mathbf{H}_m - \hat{\mathbf{H}}_m\|_F^2 + \beta \Bigl(1 - \frac{\langle \mathbf{H}_m, \hat{\mathbf{H}}_m\rangle}{\|\mathbf{H}_m\|_F \|\hat{\mathbf{H}}_m\|_F} \Bigr),
	\end{equation}
	where $N_{\text{mask}} = L_{\text{mask}} \times d_{\text{patch}}$ and $\beta$ is a weighting coefficient. The MSE term enforces element-wise reconstruction accuracy, while the cosine-similarity term emphasizes the global angular alignment between the predicted and target tensors.

	\section{Foundation Optimization Decoder}
\begin{figure*}[htbp]
    \centering
    \includegraphics[width=0.75\textwidth]{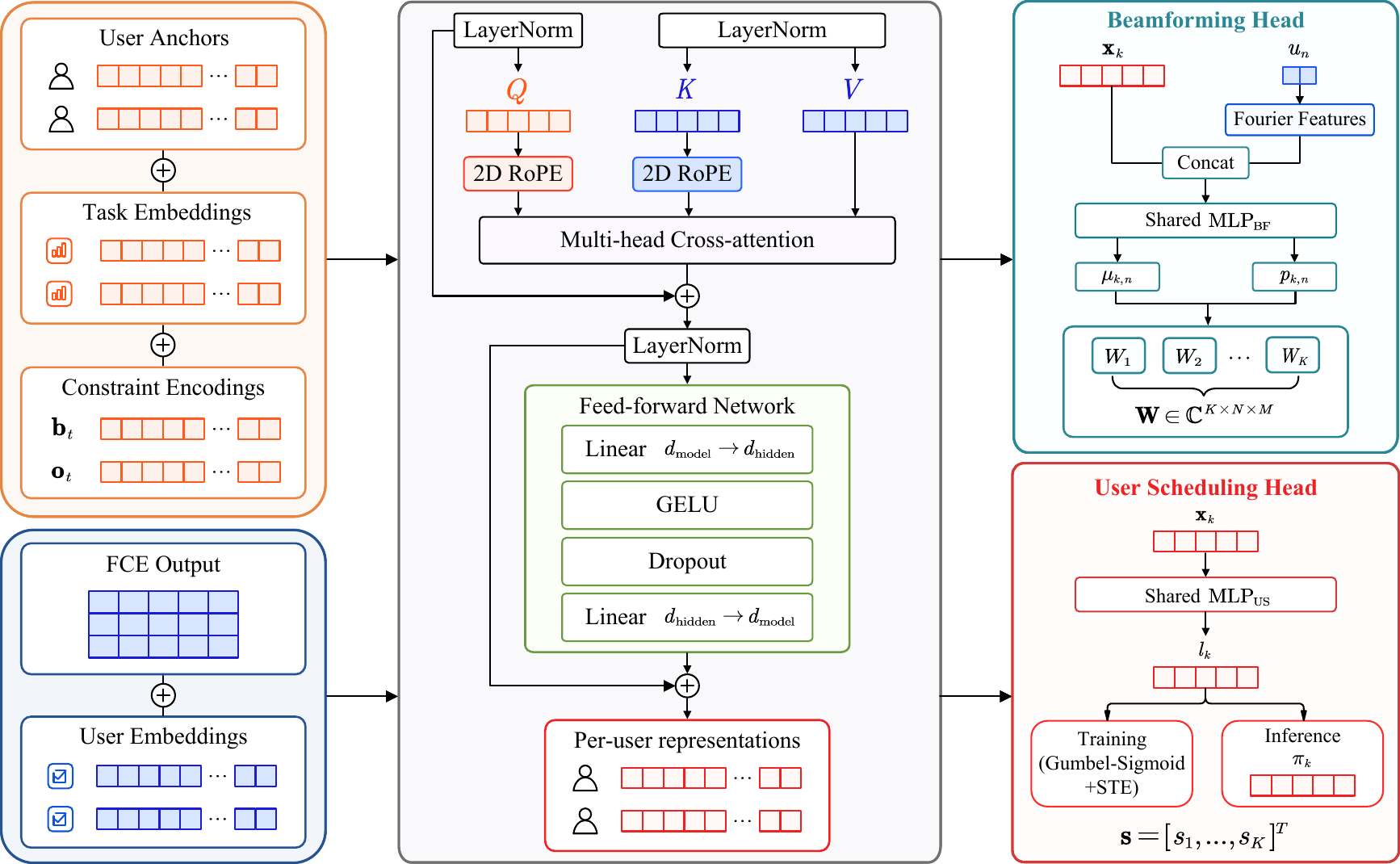}
    \caption{Network architecture of the proposed FOD.}
    \label{FOD}
	\end{figure*}

	As the downstream reasoning module, the FOD maps the universal channel representations to task-specific solutions via prompt-conditioned cross-attention, thereby enabling unified modeling of heterogeneous tasks across varying system configurations. In this section, we first detail the network architecture of the FOD, as shown in Fig.~\ref{FOD}, and then introduce a hybrid multi-task training scheme.

	\subsection{Prompt-Based Input Embedding}
	To facilitate zero-shot generalization across varying constraints and user counts, we formulate decoder input $\mathbf{X}_{\text{in}} \in \mathbb{R}^{K_{\max} \times d_{\text{model}}}$ as a composite prompt, where $K_{\max}$ denotes the maximum scheduling capacity of the BS and $d_{\text{model}}$ is the model dimension of the FOD. Inactive queries are masked during attention to decouple the input shape from varying $K$. Since optimization objectives are inherently user-centric, each query is assigned to a specific user slot and integrates three contextual components.

	\paragraph{Learnable User Anchors}
	A learnable matrix $\mathbf{A} \in \mathbb{R}^{K_{\max} \times d_{\text{model}}}$ is introduced, where each row serves as a semantic anchor for a fixed user slot. This design decouples user indexing from varying user counts, enabling the parallel generation of all user-specific outputs without architectural reconfiguration.

	\paragraph{Task Embeddings}
	Discrete task index $t$ is mapped to a vector $\mathbf{e}_t \in \mathbb{R}^{1 \times d_{\text{model}}}$ via a learnable lookup table and broadcast across all queries as $\mathbf{E}_t \in \mathbb{R}^{K_{\max} \times d_{\text{model}}}$, thereby conditioning the subsequent cross-attention on the current task.

	\paragraph{Constraint Encodings}
	System constraints delineate the boundaries of the feasible solution space. Let $\mathbf{c} \in \mathbb{R}^{K_{\max} \times 1}$ denote the per-user constraint vector with zero-padding for inactive queries. To allow distinct tasks to process identical scalar constraints along task-specific directions, we adopt a task-conditional affine encoding. For each task $t$, we learn a direction vector $\mathbf{b}_t \in \mathbb{R}^{1 \times d_{\text{model}}}$ and an offset vector $\mathbf{o}_t \in \mathbb{R}^{1 \times d_{\text{model}}}$ to yield the constraint embedding
	\begin{equation}
	\mathbf{E}_c = \mathbf{c} \mathbf{b}_t + \mathbf{1}_{K_{\max}} \mathbf{o}_t,
	\end{equation}
	where $\mathbf{b}_t$ specifies the direction along which the constraint magnitude modulates the embedding, and $\mathbf{o}_t$ defines the baseline at $\mathbf{c} = \mathbf{0}$. The linearity of this mapping facilitates robust extrapolation to constraints beyond the training range.

	The three components are summed element-wise to form the composite input embedding
	\begin{equation}
	\mathbf{X}_{\text{in}} = \mathbf{A} + \mathbf{E}_t + \mathbf{E}_c,
	\end{equation}
	where each user query jointly encodes its semantic identity, current task, and individual constraint.

	\subsection{Geometry-Aware Cross-Attention}
	To integrate prompt-based input embeddings with universal channel representations, multi-head cross-attention (MCA) is employed. Since the FCE adopts a shared coordinate matrix, patches of different users may occupy identical geometric indices. To resolve this semantic ambiguity, a learnable user identity embedding $\mathbf{u}_k \in \mathbb{R}^{1 \times d_{\text{enc}}}$ is injected into each patch of user $k$ prior to projection, yielding augmented representation $\mathbf{Z}_{\text{enc}}'$. Let $\mathbf{X}_{j-1}$ denote the input of the $j$-th block, initialized as $\mathbf{X}_0 = \mathbf{X}_{\text{in}}$. For the $i$-th head,
	\begin{equation}
	\mathbf{Q}_i = \mathbf{X}_{j-1} \mathbf{W}_Q^i, \quad \mathbf{K}_i = \mathbf{Z}_{\text{enc}}' \mathbf{W}_K^i, \quad \mathbf{V}_i = \mathbf{Z}_{\text{enc}}' \mathbf{W}_V^i,
	\end{equation}
	where $\mathbf{W}_Q^i$, $\mathbf{W}_K^i$ and $\mathbf{W}_V^i$ are learnable projection matrices.

	Beyond semantic disambiguation, user-centric queries are aligned with patch-level keys in a unified geometric space via 2D RoPE. A learnable anchor $(u_k, v_k)$ serves as the query coordinate for user $k$, whereas the patch coordinate $(u_l, v_l)$ is obtained from $\mathbf{C}_{\text{vis}}$\footnote{To enable zero-shot generalization to system dimensions, position interpolation linearly rescales the patch indices into the pre-trained range, keeping all coordinates within the interpolation regime of the RoPE basis \cite{chenExtendingContextWindow2023}.}. The rotary embedding $f_{\text{RoPE}}$ is then applied to both the query and key vectors,
	\begin{equation}
	\tilde{\mathbf{q}}_{k,i} = f_{\text{RoPE}}(\mathbf{q}_{k,i}, u_k, v_k), \quad \tilde{\mathbf{k}}_{l,i} = f_{\text{RoPE}}(\mathbf{k}_{l,i}, u_l, v_l),
	\end{equation}
	so that the RoPE-induced geometric modulation of the attention logits is governed by the relative displacement between the user anchor and each patch, implicitly biasing the model toward spatially proximate channel regions. 
	Substituting $\tilde{\mathbf{Q}}_i$, $\tilde{\mathbf{K}}_i$, and $\mathbf{V}_i$ into \eqref{head}, the per-head outputs are concatenated and linearly projected to form the MCA output. After passing through all blocks, decoded representation $\mathbf{X} \in \mathbb{R}^{K_{\max} \times d_{\text{model}}}$ is produced.

	\subsection{Multi-Task Output Heads}
	The task-specific output heads map the per-user decoded representation $\mathbf{x}_k$ to heterogeneous optimization variables.

	\subsubsection{Beamforming Head}
	The beamforming task maximizes the system sum rate by optimizing precoding tensor $\mathbf{W}$. The optimal downlink precoder admits a closed-form structure \cite{bjornsonOptimalMultiuserTransmit2014}:
	\begin{equation} \label{precoding}
	\mathbf{w}_{k,n} = \sqrt{p_{k,n}}\, \frac{\bigl(\mathbf{I}_M + \sum_{j=1}^{K} \mu_{j,n} \mathbf{h}_{j,n} \mathbf{h}_{j,n}^H\bigr)^{-1} \mathbf{h}_{k,n}}{\bigl\lVert \bigl(\mathbf{I}_M + \sum_{j=1}^{K} \mu_{j,n} \mathbf{h}_{j,n} \mathbf{h}_{j,n}^H\bigr)^{-1} \mathbf{h}_{k,n}\bigr\rVert},
	\end{equation}
	where $p_{k,n}$ is the power allocated to user $k$ on subcarrier $n$, and $\mu_{k,n}$ denotes the corresponding regularization parameter. This decomposition reduces the high-dimensional beamforming task to the joint optimization of two low-dimensional matrices: a regularization matrix $\boldsymbol{\mu}$ and a power allocation matrix $\mathbf{P}$. 

	To decouple the head from varying $N$ while preserving the frequency-domain correlations of wideband channels, we parameterize the per-subcarrier predictions as a continuous function of normalized coordinates $\{\xi_n\}_{n=1}^N \subset [0,1]$ via an implicit neural representation (INR) \cite{tancikFourierFeaturesLet2020}. 
	However, directly feeding $\xi_n$ into a multi-layer perceptron (MLP) incurs a spectral bias toward low-frequency functions. Therefore, each coordinate is lifted into a high-frequency feature space via a learnable frequency vector $\mathbf{b}$
	\begin{equation}
	\phi(\xi_n) = \left[\sin(2\pi \xi_n \mathbf{b}),\, \cos(2\pi \xi_n \mathbf{b})\right].
	\end{equation}
	For each user $k$ and subcarrier $n$, the head concatenates the user latent state with the Fourier embedding to produce two pre-activation scalars
	\begin{equation}
	(\tilde{\mu}_{k,n}, \tilde{p}_{k,n}) = \operatorname{MLP}_{\text{BF}} \left(\operatorname{Concat}(\mathbf{x}_k, \phi(\xi_n))\right).
	\end{equation}
	The non-negativity of regularization parameters is enforced by $\mu_{k,n} = \operatorname{softplus}(\tilde{\mu}_{k,n})$ and the total power constraint is imposed via a differentiable simplex projection $\mathbf{P} = \operatorname{Proj}_{\mathcal{S}}(\tilde{\mathbf{P}})$, where $\mathcal{S}$ is the feasible power set and $\tilde{\mathbf{P}}$ denotes the power pre-activation matrix. To preclude power leakage, pre-activations on inactive user slots are masked prior to projection.

	\subsubsection{User Scheduling Head}
	The user scheduling task is formulated as a binary subset selection problem over the candidate users. The output consists of $K$ binary decisions and is independent of system dimension $(N, M)$. Since MCA has already embedded the multi-user context into latent state $\mathbf{x}_k$, the scheduling head requires no additional inter-user interaction module and employs a shared per-user MLP that maps each $\mathbf{x}_k$ to a selection logit $l_k$ and its probability $\pi_k$:
	\begin{equation}
	l_k = \operatorname{MLP}_{\text{US}}(\mathbf{x}_k), \quad \pi_k = \operatorname{sigmoid}(l_k).
	\end{equation}
	To enable end-to-end backpropagation through discrete variable $s_k$, we adopt the Gumbel-Sigmoid relaxation \cite{jangCategoricalReparameterizationGumbelSoftmax2017} during training. Perturbing each logit with additive logistic noise yields the soft selection variable
	\begin{equation}
	\tilde{s}_k = \operatorname{sigmoid} \biggl( \frac{l_k + \ell_k}{\tau} \biggr), \quad \ell_k = \log U_k - \log(1 - U_k),
	\end{equation}
	where $U_k \sim \mathcal{U}(0,1)$, $\ell_k$ follows a standard logistic distribution, and the temperature $\tau>0$ is annealed over training to tighten $\tilde{s}_k$ toward its discrete limit.
	To enable binary inputs in the sum-rate evaluation while preserving gradient flow, we apply the straight-through estimator (STE) \cite{bengioEstimatingPropagatingGradients2013}:
	\begin{equation}
	s_k = \mathbb{I}(\tilde{s}_k > 0.5) - \operatorname{sg}(\tilde{s}_k) + \tilde{s}_k,
	\end{equation}
	where $\mathbb{I}(\cdot)$ is the indicator function and $\operatorname{sg}(\cdot)$ denotes the stop-gradient operator. The forward pass yields hard decisions consistent with the physical interference model while the backward pass propagates surrogate gradients through $\tilde{s}_k$. At inference, stochastic sampling is replaced by deterministic thresholding $s_k = \mathbb{I}(\pi_k > 0.5)$, and the per-user decisions are stacked into the global scheduling vector $\mathbf{s} = [s_1, \ldots, s_K]^T$.

	\subsection{Hybrid Multi-Task Training}
	Training the FOD for multi-task optimization faces three primary challenges: the heterogeneity of task objectives, limited generalization under varying constraints, and the lack of optimal labels. To address these challenges, we propose a hybrid training strategy comprising three complementary components: 
	\begin{itemize}
		\item an interleaved mini-batch protocol that mitigates conflicting task gradients;
		\item a dynamic constraint sampling scheme that exposes the decoder to a continuum of constraint settings; 
		\item a two-stage curriculum that transitions from supervised warm-up to unsupervised fine-tuning.
	\end{itemize}

	To coordinate heterogeneous objectives, we adopt an interleaved mini-batch protocol. Specifically, we sample a task index at each iteration and construct the mini-batch from the corresponding data distribution, so that each parameter update is driven solely by the gradient of the active task. This temporal decoupling prevents tasks with large-magnitude losses from dominating the updates, thereby promoting balanced convergence across tasks.

	To improve generalization over the feasible solution space, we further introduce a dynamic constraint sampling scheme. At each iteration, user-specific constraints are drawn from a continuous uniform distribution and serve a dual role: they parameterize the input prompts and define the corresponding thresholds in the loss functions. This scheme enables the FOD to learn a generalized mapping from dynamic constraints to constraint-conditioned policies. 
	
	Building upon these two training strategies, we present the two-stage curriculum.
	\subsubsection{Supervised Warm-Up}
	To stabilize early-stage optimization, we begin with a supervised warm-up phase, thereby providing a well-conditioned initialization for the subsequent unsupervised fine-tuning.

	\paragraph{Beamforming}
	Since the beamforming head predicts $(\boldsymbol{\mu}, \mathbf{P})$ rather than the precoding tensor directly, we supervise these two parameter matrices using targets obtained by decomposing the WMMSE solution. Because $\boldsymbol{\mu}$ and $\mathbf{P}$ differ substantially in scale, we adopt an NMSE criterion to provide scale-normalized supervision:
	\begin{equation}
	\mathcal{L}_{\text{BF-SL}} = \frac{\|\boldsymbol{\mu} - \boldsymbol{\mu}^*\|_F^2}{\|\boldsymbol{\mu}^*\|_F^2} + \frac{\|\mathbf{P} - \mathbf{P}^*\|_F^2}{\|\mathbf{P}^*\|_F^2},
	\end{equation}
	where $\boldsymbol{\mu}^{*} \in \mathbb{R}^{K \times N}$ and $\mathbf{P}^{*} \in \mathbb{R}^{K \times N}$ denote the target regularization matrix and power allocation matrix, respectively.

	\paragraph{User Scheduling}
	For the discrete scheduling task, we adopt label smoothing \cite{szegedyRethinkingInceptionArchitecture2016} to mitigate overconfident sigmoid predictions and define an MSE loss against the smoothed algorithmic labels:
	\begin{equation}
	\mathcal{L}_{\text{US-SL}} = \frac{1}{K} \sum_{k=1}^{K} \left(\pi_k - \tilde{s}_k^{*}\right)^2,
	\end{equation}
	where $\tilde{s}_k^*$ denotes the smoothed target of the binary label generated by the greedy scheduling algorithm. By keeping the targets away from the hard limits, label smoothing mitigates sigmoid saturation, thereby facilitating the transition to the Gumbel-Sigmoid relaxation in the unsupervised stage.

	\subsubsection{Unsupervised Fine-Tuning}
	To surpass the performance ceiling imposed by algorithmic supervision, we fine-tune the FOD with unsupervised objectives that directly optimize the system utility.

	\paragraph{Beamforming}
	To maximize the system sum rate, the unsupervised beamforming loss is formulated as
	\begin{equation}
	\mathcal{L}_{\text{BF-UL}} = -\frac{1}{KN} \sum_{k=1}^{K} \sum_{n=1}^{N} \log_2(1 + \gamma_{k,n}).
	\end{equation}
	The normalization converts the objective into the average per-user per-subcarrier rate, making the loss magnitude comparable across different dimensions $(K, N)$.

	\paragraph{User Scheduling}
	To promote per-user QoS satisfaction, we adopt an augmented Lagrangian formulation and define the unsupervised loss as
	\begin{equation}
	\mathcal{L}_{\text{US-UL}}=\frac{1}{KN}\biggl(-\sum_{k=1}^{K}R_k(\mathbf{s})+\sum_{k=1}^{K}\lambda_k C_k+\frac{\rho}{2}\sum_{k=1}^{K}C_k^2\biggr),
	\end{equation}
	where $C_k = s_k \operatorname{ReLU}(R_{\min} - R_k(\mathbf{s}))$, $\lambda_k$ denotes the dual variable, and $\rho$ is the penalty coefficient. The dual variables are updated once per epoch via projected dual ascent
	\begin{equation}
	\lambda_k^{e+1} = \operatorname{clip} \left(\lambda_k^e + \rho \bar{c}_k, 0, \lambda_{\max}\right),
	\end{equation}
	where $\bar{c}_k = \mathbb{E}[s_k(R_{\min} - R_k(\mathbf{s}))]$ denotes the epoch-averaged signed residual, and $\lambda_{\max}$ caps the dual variables to avoid unbounded growth. When the QoS constraints are satisfied with slack, the signed residual becomes negative, thereby enabling the policy to explore more aggressive scheduling decisions.

	\section{Modular Task Adaptation}
	The modular WFM architecture enables efficient adaptation to communication tasks beyond the pre-training objectives. We present two adaptation pathways: upstream FCE adaptation for channel estimation via pilot-aware input reformulation, and downstream FOD adaptation for beam selection via parameter-efficient fine-tuning.

	\subsection{Upstream Adaptation: Channel Estimation}
	To assess the representational fidelity of the FCE, we recast pilot-based channel estimation as a masked reconstruction problem that recovers full CSI from the sparse LS estimate.

	\subsubsection{Adaptation Architecture}
	Since the patch-embedding layer of the FCE is designed for fully observed CSI, it is replaced with a pilot-aware input adapter. For each user $k$, the adapter forms a three-channel input by stacking the real and imaginary parts of $\mathbf{H}_{\text{LS}}^k$ with a binary pilot mask $\mathbf{p} \in \{0, 1\}^{N}$ broadcast along the antenna dimension. This input is then projected into the encoder dimension through a learnable 2D convolution, and a noise-level embedding is added to each patch token, thereby conditioning the reconstruction on the operating SNR.

	Unlike the joint multi-user encoding used in FCE pre-training, channel estimation is performed on a per-user basis. The encoder processes the per-user patch sequence without masking, allowing zero-filled non-pilot positions to serve as contextual placeholders. A pilot-aware partitioning scheme then reconstructs the input of the decoder: patches containing at least one pilot subcarrier are forwarded to the decoder while fully unobserved patches are replaced by learnable mask tokens. The decoder is initialized from the pre-trained weights and reconstructs all patches, with the LS estimate added residually to the output:
	\begin{equation}
	\hat{\mathbf{H}} = \operatorname{Unpatch}\bigl(f_{\text{pred}}(\mathbf{Z}_{\text{dec}})\bigr) + \mathbf{H}_{\text{LS}},
	\end{equation}
	where $\mathbf{Z}_{\text{dec}}$ denotes the decoder output representation, $f_{\text{pred}}(\cdot)$ is the linear prediction head, and $\operatorname{Unpatch}(\cdot)$ maps the patch predictions to the original CSI dimensions. This formulation allows the network to learn a residual correction over the LS estimate, facilitating convergence and training stability. During adaptation, the pre-trained encoder remains frozen and only the input adapter, the decoder, and the prediction head are updated, keeping the encoding pathway shared with the downstream tasks intact.

	\subsubsection{Training Objective}
	Since the NMSE does not explicitly enforce phase coherence, we train the model with a composite loss that augments the NMSE with a phase-alignment term
	\begin{equation}
	\mathcal{L}_{\text{CE}} = \frac{\|\mathbf{H} - \hat{\mathbf{H}}\|_F^2}{\|\mathbf{H}\|_F^2} + \lambda \biggl(1-\frac{\bigl|\langle \mathbf{H}, \hat{\mathbf{H}}\rangle\bigr|}{\|\mathbf{H}\|_F\,\|\hat{\mathbf{H}}\|_F}\biggr),
	\end{equation}
	where $\lambda$ controls the relative weight of the phase-alignment term. The loss is evaluated over both pilot and non-pilot positions, while the pilot ratio $r_p$ and SNR are resampled per batch to expose the model to diverse observation conditions.

	\subsection{Downstream Adaptation: Beam Selection}
	To assess the cross-task adaptability of the FOD, we adapt it to beam selection, which is structurally distinct from the pre-training tasks. 

	\subsubsection{Adaptation Architecture}
	To prevent catastrophic forgetting and reduce trainable parameter overhead, the FCE encoder and the FOD backbone remain frozen while adaptation is confined to a small set of newly introduced parameters. At the input stage, a new entry is appended to the shared task-embedding lookup table, and a corresponding pair of direction and offset vectors is added to the constraint encoder. At the output stage, a task-specific head is appended to map decoder latent representations to per-user selection logits over the codebook.

	\subsubsection{Beam Selection Head}
	To provide an initialization for selection, we construct a single-user gain prior $\bar{\mathbf{g}}_k\in\mathbb{R}^{|\mathcal{C}|}$ by applying per-user normalization to the DFT-codeword gains computed from the subcarrier-averaged channel. The output head then augments this prior with a learnable multi-user residual correction $\boldsymbol{\delta}_k$ to produce the per-user selection logits
	\begin{equation}
	\boldsymbol{\ell}_k = \bar{\mathbf{g}}_k + \eta\,\boldsymbol{\delta}_k,
	\end{equation}
	where $\eta$ scales the residual correction relative to the gain prior.

	To decouple the number of trainable parameters from $|\mathcal{C}|$, we parameterize $\boldsymbol{\delta}_k$ in a low-dimensional Fourier angular basis. A shared feature extractor followed by a linear projection maps the latent state $\mathbf{x}_k$ into the Fourier coefficients $\mathbf{a}_k \in \mathbb{R}^{2L}$. Basis matrix $\boldsymbol{\Phi}\in\mathbb{R}^{|\mathcal{C}|\times 2L}$ evaluates the first $L$ angular harmonics at the codebook angles $\theta_i=2\pi i/|\mathcal{C}|$. The row associated with codeword $i$ is given by
	\begin{equation}
	\boldsymbol{\Phi}_{i} = \frac{1}{\sqrt{L}}\!\left[\cos(\theta_i),\sin(\theta_i),\ldots,\cos(L\theta_i),\sin(L\theta_i)\right],
	\end{equation}
	so that $\boldsymbol{\delta}_k=\boldsymbol{\Phi}\mathbf{a}_k$. 
	Zero-initializing the linear projection makes the output reduce to the gain prior at initialization, providing a warm start for residual learning.

	Since the formulation does not enforce codeword exclusivity among users, we apply the Hungarian algorithm \cite{kuhnHungarianMethodAssignment1955} as a post-processing step at inference when $|\mathcal{C}|\ge K$, yielding a one-to-one user-codeword assignment.

	\subsubsection{Training Objective}
	To enable gradient flow through the discrete selection, we apply the Gumbel-Softmax relaxation with temperature annealing and the STE to obtain differentiable surrogates for the per-user selection vectors. The model is trained in an unsupervised manner by minimizing the normalized loss function:
	\begin{equation}
	\mathcal{L}_{\text{BS-UL}} = -\frac{1}{KN}\sum_{k=1}^{K}\sum_{n=1}^{N}\log_2\!\bigl(1+\gamma_{k,n}(\mathbf{i})\bigr),
	\end{equation}
	where $\gamma_{k,n}(\mathbf{i})$ is evaluated with the straight-through selection vectors during training and the Hungarian-assigned codewords at inference.

	\section{Numerical Results}
	In this section, we evaluate the proposed WFM on four tasks to validate the fidelity of the learned channel representations and the effectiveness of modular task adaptation. For each task, we examine training convergence across random seeds and evaluate task performance under varying constraints. We then assess zero-shot generalization to unseen propagation environments and system configurations, and finally compare inference latency with task-specific baselines.

	\subsection{Simulation Setup and Datasets}
	\begin{figure*}[tbp]
		\centering
		\includegraphics[width=0.75\textwidth]{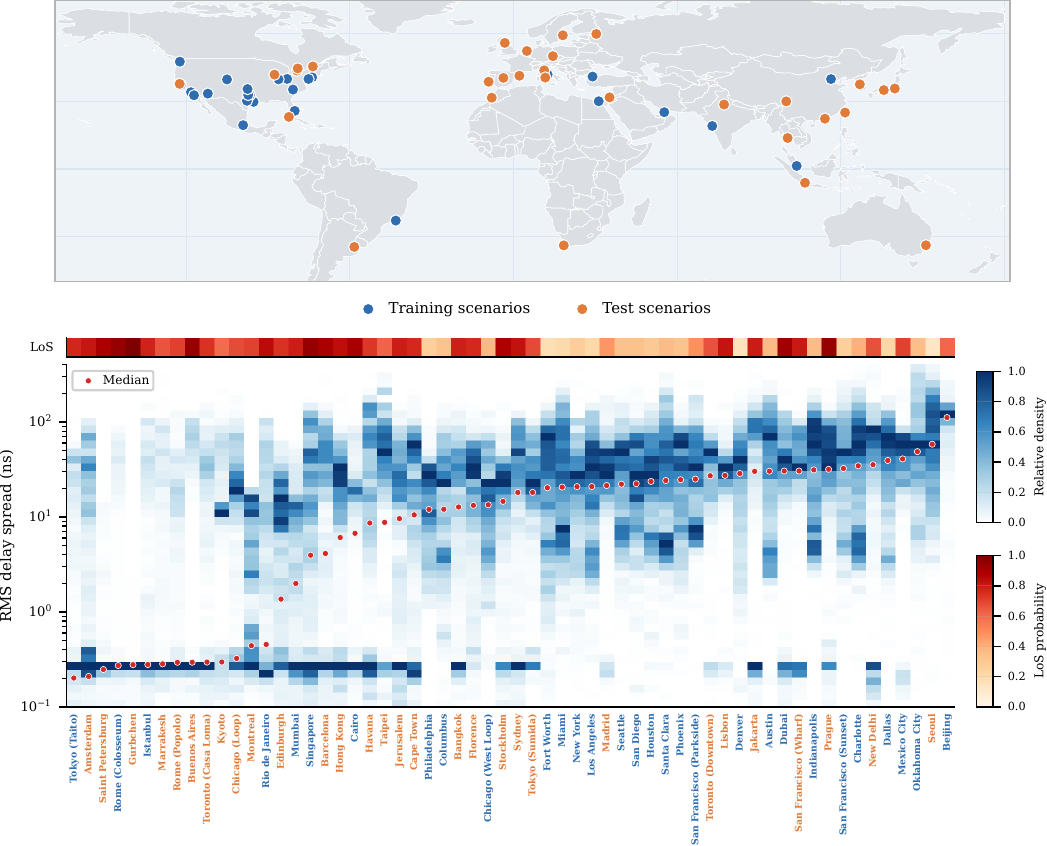}
		\caption{Overview of the multi-scenario CSI corpus. The upper panel maps the geographic distribution of the urban scenarios, while the lower panel shows the per-scenario distributions of RMS delay spread as a column-normalized density heatmap, with the strip above it encoding the LoS probability.}
		\label{fig:corpus}
	\end{figure*}

	Channel realizations are drawn from the DeepMIMO dataset \cite{alkhateebDeepMIMOGenericDeep2019}, which leverages the Wireless InSite ray-tracer to model site-specific propagation. As shown in Fig.~\ref{fig:corpus}, we curate a corpus of 60 urban scenarios, split into 30 scenarios for training and in-domain validation and 30 for cross-scenario generalization. The corpus exhibits wide variations in RMS delay spread and line-of-sight (LoS) probability, exposing the model to heterogeneous propagation conditions. To assess adaptability across system configurations, we define a seen system dimension grid ($K \in \{4, 8, 12\}$, $N \in \{32, 64, 128\}$, $M \in \{8, 16, 32\}$) and an unseen grid ($K \in \{6, 11, 16\}$, $N \in \{48, 96, 192\}$, $M \in \{12, 24, 48\}$), each spanning 27 distinct $(K, N, M)$ combinations. Within the unseen grid, two values along each dimension fall within the seen range while the third extends beyond it, which covers both interpolation and extrapolation.

	Combining the scenario sets with the configuration grids yields one in-domain set and three zero-shot evaluation sets. The in-domain set (seen scenarios, seen configurations) contains 5000 CSI samples per scenario, of which 4500 and 500 are used for training and validation, respectively. Each zero-shot evaluation set contains 500 CSI samples per scenario:
	\begin{itemize}
		\item the cross-scenario set (unseen scenarios, seen configurations); 
		\item the cross-configuration set (seen scenarios, unseen configurations);
		\item the joint set (unseen scenarios, unseen configurations). 
	\end{itemize}

	For the hyperparameter settings, the FCE comprises a 12-layer encoder ($d_{\text{enc}}=512$, 8 heads) and a 4-layer decoder ($d_{\text{dec}}=256$, 8 heads), with a patch size of $(n,m)=(4,2)$ and a masking ratio of 75\%. The FOD adopts a 6-layer MCA decoder ($d_{\text{model}}=512$, 8 heads) with $K_{\max}=16$ user slots. All trainable modules are optimized using AdamW with a cosine learning rate schedule and linear warm-up. 

	To evaluate the WFM across different tasks, we compare it with the following task-specific baselines, including numerical algorithms and learning-based methods:
	\begin{itemize}
		\item \textbf{LS:} LS channel estimation on pilot subcarriers, followed by linear interpolation over non-pilot subcarriers.
		\item \textbf{LMMSE:} LMMSE channel estimation using the oracle frequency-domain correlation matrix for joint denoising and interpolation \cite{vandebeekChannelEstimationOFDM1995}.
		\item \textbf{WMMSE:} A precoding algorithm for weighted sum-rate maximization that alternates among receive filters, MSE weights, and transmit precoders \cite{shiIterativelyWeightedMMSE2011}.
		\item \textbf{Greedy Scheduling:} Sequentially selects the user that yields the largest marginal sum-rate gain while satisfying per-user QoS constraints.
		\item \textbf{DFT Greedy:} Each user independently selects the DFT codeword that maximizes its single-user channel gain, without accounting for inter-user interference.
		\item \textbf{Seq Greedy:} Sequentially assigns each user the codeword that maximizes the sum rate given the codewords already assigned to previous users.
		\item \textbf{Transformer:} A dimension-agnostic encoder-only Transformer trained from scratch for each task, using the same input-output formulation and task objective as the WFM.
		\item \textbf{WLLM:} A baseline that replaces the FOD decoder with a pre-trained GPT-2 backbone adapted with LoRA, while keeping the remaining WFM pipeline unchanged.
	\end{itemize}

	\subsection{Pre-Training and Channel Estimation}
	\begin{figure}[htbp]
		\centering
		\includegraphics[width=0.43\textwidth]{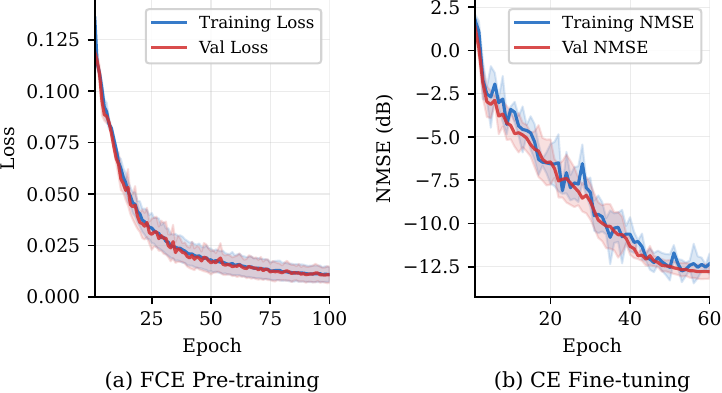}
		\caption{Training convergence curves for (a) FCE pre-training and (b) channel estimation fine-tuning. Shaded regions indicate standard deviations over three random seeds.}
		\label{fig5}
	\end{figure}

	Fig.~\ref{fig5} illustrates the convergence behavior of FCE pre-training and channel estimation fine-tuning. In Fig.~\ref{fig5}(a), the FCE loss decreases steadily, with closely matched training and validation curves and narrow shaded bands, indicating stable convergence across random seeds. In Fig.~\ref{fig5}(b), fine-tuning for channel estimation reduces the validation NMSE by more than 10 dB, demonstrating the fidelity of the learned channel representations. 

	\begin{figure*}[htbp]
		\centering
		\includegraphics[width=0.80\textwidth]{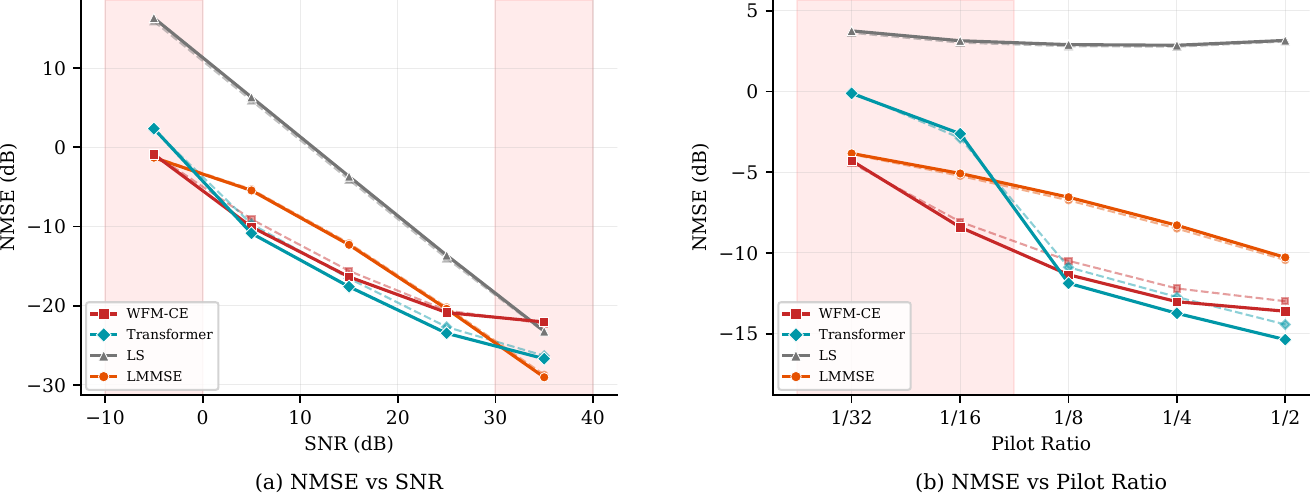}
		\caption{Channel estimation performance of WFM-CE against baseline methods: (a) NMSE versus SNR and (b) NMSE versus pilot ratio. Solid and dashed curves denote results on the in-domain and cross-scenario sets, respectively, while pink shaded regions indicate OOD regimes outside the training range.}
		\label{fig6}
	\end{figure*}

	Fig.~\ref{fig6} evaluates channel estimation performance across varying SNRs and pilot ratios. In Fig.~\ref{fig6}(a), WFM-CE achieves its largest gain over LMMSE in the moderate-SNR regime, even though LMMSE exploits oracle channel statistics. This advantage diminishes at both extremes: in the low-SNR out-of-distribution (OOD) regime, the two methods perform comparably; at high SNRs, the residual error becomes increasingly dominated by interpolation rather than noise, allowing LMMSE to recover an advantage over WFM-CE. In Fig.~\ref{fig6}(b), LS is largely insensitive to pilot density, whereas WFM-CE and LMMSE both improve with denser pilots. Moreover, WFM-CE consistently outperforms LMMSE, with the gap narrowing at the sparsest pilot setting. 
	Across both subfigures, WFM-CE and Transformer perform comparably within the training range, but Transformer degrades markedly in the low-SNR and sparse-pilot OOD regimes, indicating that WFM-CE generalizes more reliably to unseen system parameters. The close alignment between the in-domain and cross-scenario WFM-CE curves demonstrates robust generalization to unseen propagation scenarios.

	\subsection{Multi-Task Optimization}
	\begin{figure*}[htbp]
		\centering
		\includegraphics[width=\textwidth]{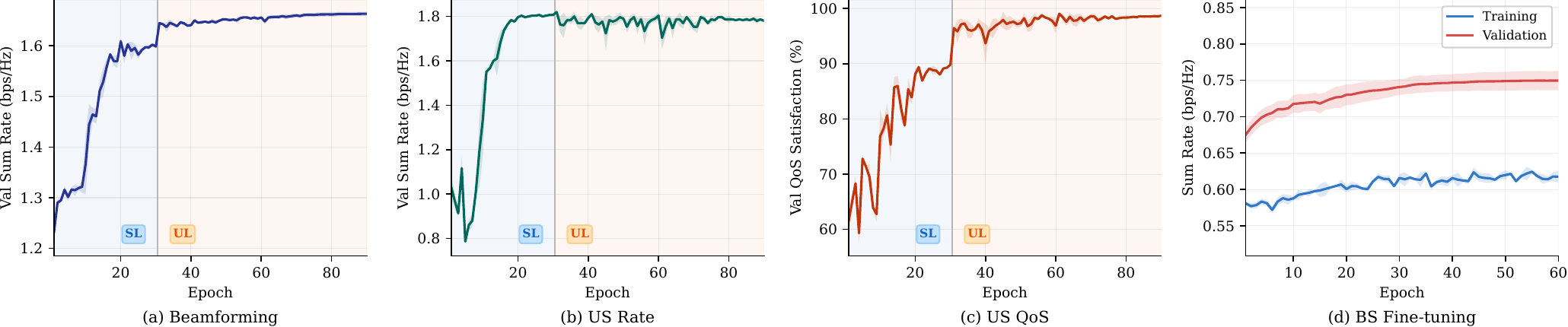}
		\caption{Training convergence curves for (a) FOD beamforming, (b) user scheduling rate, (c) user scheduling QoS satisfaction, and (d) beam selection fine-tuning. Blue and pink backgrounds in (a)-(c) denote the SL and UL training phases while shaded regions indicate standard deviations over three random seeds.}
		\label{fig7}
	\end{figure*}

	Fig.~\ref{fig7} shows the convergence behavior of FOD pre-training and beam selection fine-tuning. In Fig.~\ref{fig7}(a), the sum rate increases rapidly during supervised learning (SL) and continues to improve during unsupervised learning (UL), indicating that UL provides additional gains beyond SL. In Fig.~\ref{fig7}(b), the sum rate improves quickly during SL but saturates shortly after the transition to UL, implying that SL already captures most of the sum-rate gain. In Fig.~\ref{fig7}(c), the QoS satisfaction improves throughout both phases, highlighting the role of UL in enhancing QoS satisfaction. In Fig.~\ref{fig7}(d), fine-tuning for beam selection converges rapidly, and the validation rate remains higher than the training rate due to inference-only Hungarian post-processing, which resolves beam collisions not prevented by Gumbel-Softmax sampling during training.

	\begin{figure*}[htbp]
		\centering
		\includegraphics[width=\textwidth]{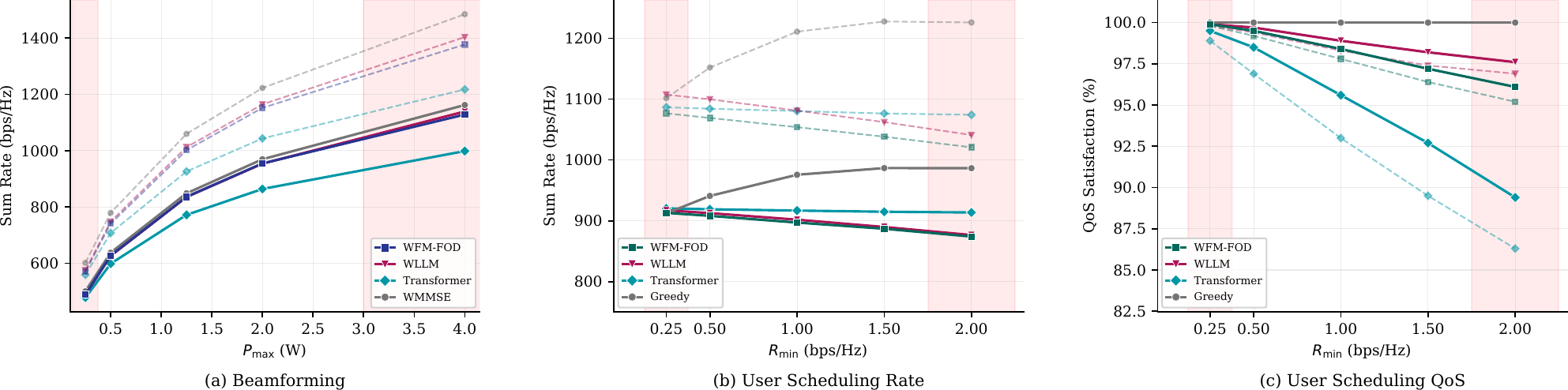}
		\caption{Performance of WFM-FOD against baseline methods: (a) beamforming sum rate versus $P_{\max}$, (b) user scheduling sum rate versus $R_{\min}$, and (c) user scheduling QoS satisfaction versus $R_{\min}$. Solid and dashed curves denote results on the in-domain and cross-scenario sets, respectively, while pink shaded regions indicate OOD regimes outside the training range.}
		\label{fig8}
	\end{figure*}

	Fig.~\ref{fig8} evaluates FOD performance under varying constraint parameters. In Fig.~\ref{fig8}(a), WFM-FOD and WLLM closely track WMMSE across the entire power range on both the in-domain and cross-scenario sets, whereas Transformer consistently underperforms. Fig.~\ref{fig8}(b) and (c) jointly reveal a rate-QoS trade-off as $R_{\min}$ tightens: WFM-FOD and WLLM progressively trade off sum-rate performance to maintain high QoS satisfaction, whereas Transformer holds the sum rate nearly constant at the cost of a pronounced QoS degradation. This contrast indicates that WFM-FOD reliably enforces the per-user QoS constraint, whereas Transformer fails to adapt as the constraint tightens.

	\begin{figure}[htbp]
		\centering
		\includegraphics[width=0.42\textwidth]{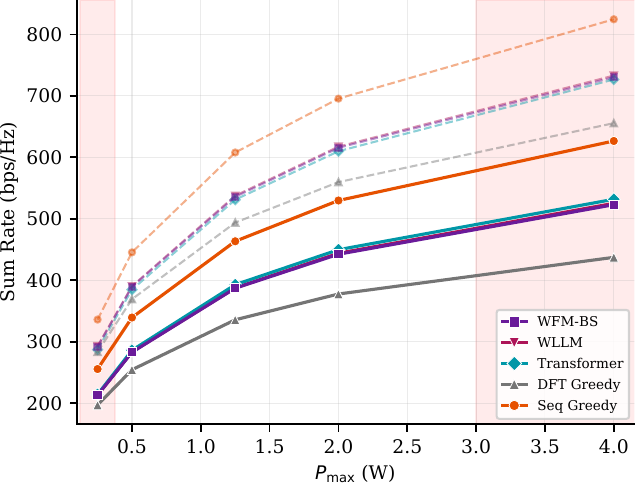}
		\caption{Beam selection performance of WFM-BS against baseline methods versus $P_{\max}$. Solid and dashed curves denote results on the in-domain and cross-scenario sets, respectively, while pink shaded regions indicate OOD regimes outside the training range.}
		\label{fig9}
	\end{figure}

	Fig.~\ref{fig9} evaluates beam selection performance versus $P_{\max}$. Across the entire power range, WFM-BS performs on par with WLLM and closely tracks Transformer, while all three learning-based methods substantially outperform DFT Greedy, indicating that the learned representations effectively capture the multi-user interference structure. Although Seq Greedy attains the highest sum rate, WFM-BS maintains a moderate performance gap while requiring only a single forward pass rather than a sequential search. Furthermore, the consistent performance trend of WFM-BS on the cross-scenario set demonstrates robust generalization to unseen scenarios and power constraints.

	\subsection{Generalizability}
	\begin{figure*}[htbp]
		\centering
		\includegraphics[width=\textwidth]{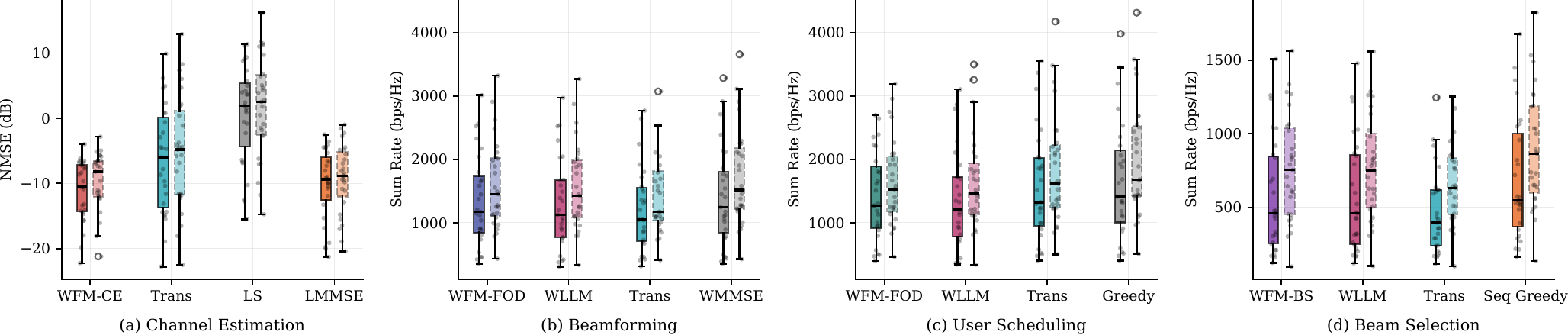}
		\caption{Generalizability of the WFM against corresponding baselines across four tasks: (a) channel estimation, (b) beamforming, (c) user scheduling, and (d) beam selection. Dark and light boxes denote results on the cross-configuration and joint sets, respectively.}
		\label{fig10}
	\end{figure*}

	Fig.~\ref{fig10} evaluates zero-shot generalization across 27 unseen $(K, N, M)$ configurations. In Fig.~\ref{fig10}(a), WFM-CE outperforms LS and Transformer while remaining comparable to LMMSE. In Fig.~\ref{fig10}(b) and (c), WFM-FOD remains competitive with the WMMSE and Greedy baselines, respectively, while matching WLLM and surpassing Transformer. In Fig.~\ref{fig10}(d), WFM-BS remains on par with WLLM and consistently outperforms Transformer, while narrowing the performance gap to the Seq Greedy baseline. Across all four tasks, the WFM exhibits only minor shifts in both median and dispersion between the cross-configuration and joint sets, indicating robust generalization to unseen system configurations and propagation environments.

	\subsection{Computational Complexity and Latency}
	\begin{table*}[t]
		\centering
		\caption{Computational Complexity, Inference Latency, and Performance Comparison}
		\label{tab:latency}
		\renewcommand{\arraystretch}{1.25}
		\setlength{\tabcolsep}{4pt}
		\scriptsize
		\begin{tabular}{@{}cc ccc c @{\hspace{12pt}} cc ccc c@{}}
			\toprule
			Task & Method & Params (M) & FLOPs (G) & Latency (ms) & Perf. &
			Task & Method & Params (M) & FLOPs (G) & Latency (ms) & Perf. \\
			\midrule
			\multirow{4}{*}{\shortstack{Channel\\Estimation}}
			& \textbf{WFM-CE} & \textbf{41.37} & \textbf{378.24} & \textbf{1.87} & \textbf{-13.3 dB} &
			\multirow{4}{*}{Beamforming}
			& \textbf{WFM-FOD} & \textbf{19.59} & \textbf{31.51} & \textbf{2.67} & \textbf{98.5\%} \\
			& Transformer & 2.71 & 24.53 & 0.86 & -14.0 dB &
			& Transformer & 1.21 & 7.30 & 0.52 & 90.8\% \\
			& LS    & -- & -- & 0.22 & 1.50 dB &
			& WLLM  & 128.66 & 52.46 & 3.82 & 99.0\% \\
			& LMMSE & -- & -- & 0.39 & -8.98 dB &
			& WMMSE & -- & -- & 104 & 100\% \\
			\midrule
			\multirow{4}{*}{\shortstack{User\\Scheduling}}
			& \textbf{WFM-FOD} & \textbf{19.19} & \textbf{29.50} & \textbf{2.22} & \textbf{93.2 / 98.7\%} &
			\multirow{4}{*}{\shortstack{Beam\\Selection}}
			& \textbf{WFM-BS} & \textbf{19.08} & \textbf{29.50} & \textbf{0.95} & \textbf{83.6\%} \\
			& Transformer & 1.04 & 5.47 & 0.43 & 94.3 / 96.2\% &
			& Transformer & 1.06 & 5.47 & 0.45 & 84.8\% \\
			& WLLM  & 128.27 & 50.45 & 3.37 & 94.1 / 99.1\% &
			& WLLM  & 128.15 & 50.45 & 2.17 & 83.7\% \\
			& Greedy & -- & -- & 30.4 & 100 / 100\% &
			& Seq Greedy & -- & -- & 2.15 & 100\% \\
			\bottomrule
		\end{tabular}

		\vspace{3pt}
		\begin{flushleft}
		\footnotesize
		\textit{Note:} The performance metric is NMSE for channel estimation, the normalized sum rate for beamforming and beam selection, and the normalized sum rate / QoS satisfaction for user scheduling.
		\end{flushleft}
	\end{table*}

	Table~\ref{tab:latency} reports the computational complexity, inference latency and performance of the WFM and the corresponding baselines across four tasks. Since the FCE is shared across downstream tasks, the reported parameters, FLOPs and latencies of WFM-FOD and WFM-BS exclude the FCE and account only for the FOD backbone and task-specific heads. For channel estimation, WFM-CE incurs higher latency than LS and LMMSE due to the FCE encoder-decoder overhead, but achieves substantially lower NMSE without requiring oracle channel statistics. For the pre-training tasks, WFM-FOD runs approximately $39\times$ faster than WMMSE and $14\times$ faster than Greedy Scheduling, while incurring only a marginal performance loss. For beam selection, WFM-BS is $2.3\times$ faster than Seq Greedy with a moderate performance gap. Across all tasks, the WFM achieves performance on par with WLLM while requiring a substantially smaller parameter count and incurring lower FLOPs and latency. 
	Although task-specific Transformer is more compact, it must be trained and deployed separately for each task, while the WFM supports all four tasks with a single shared backbone, thereby achieving a favorable complexity-latency-performance trade-off.

	\subsection{Ablation Study}
	\begin{table*}[t]
		\centering
		\caption{Ablation Study of the Proposed WFM}
		\label{tab:ablation}
		\renewcommand{\arraystretch}{1.3}
		\setlength{\tabcolsep}{6pt}
		\scriptsize
		\begin{tabular}{|l|ccc|ccc|ccc|}
			\hline
			\multirow{2}{*}{Variant}
			& \multicolumn{3}{c|}{BF Rate (\%) $\uparrow$}
			& \multicolumn{3}{c|}{US Rate (\%) $\uparrow$}
			& \multicolumn{3}{c|}{US QoS (\%)  $\uparrow$} \\
			\cline{2-10}
			& ID & Constr & Config & ID & Constr & Config & ID & Constr & Config \\
			\hline
			\textbf{Full WFM} & \textbf{98.5} & \textbf{97.4} & \textbf{96.6} & \textbf{93.2} & \textbf{94.4} & \textbf{88.3} & \textbf{98.7} & \textbf{98.0} & \textbf{98.4} \\
			\hline
			w/o FOD multi-task training & 99.6 & 98.7 & 97.2 & 92.4 & 91.9 & 87.4 & 97.1 & 96.7 & 97.0 \\
			w/o FOD supervised warm-up & 74.7 & 71.8 & 69.7 & 44.4 & 44.8 & 34.0 & 58.6 & 53.6 & 52.3 \\
			w/o FOD unsupervised fine-tuning & 94.8 & 92.9 & 92.8 & 96.3 & 96.5 & 92.6 & 95.3 & 95.5 & 89.1 \\
			w/o Affine constraint encoding & 98.4 & 92.8 & 96.3 & 92.9 & 92.4 & 87.6 & 98.9 & 98.7 & 98.7 \\
			w/o Geometry-aware cross-attention & 76.0 & 73.6 & 71.1 & 42.7 & 42.9 & 32.7 & 58.7 & 53.1 & 53.3 \\
			\hline
		\end{tabular}

		\vspace{3pt}
		\begin{flushleft}
		\footnotesize
		\textit{Note:} BF and US denote beamforming and user scheduling, respectively. ID denotes the in-domain set, while Constr and Config denote the constraint and configuration OOD regimes on the seen scenarios, respectively. BF Rate and US Rate are normalized by the corresponding numerical baselines.
		\end{flushleft}
	\end{table*}

	Table~\ref{tab:ablation} presents an ablation study of the proposed WFM. Removing geometry-aware cross-attention or supervised warm-up causes severe degradation across all regimes, confirming the indispensability of the former for query-to-representation alignment and the necessity of the latter for stable convergence. Omitting unsupervised fine-tuning increases the US rate but reduces the BF rate and QoS satisfaction, indicating that this phase improves upon the supervised beamforming solution while trading scheduling rate for QoS enforcement. Removing affine constraint encoding leaves in-domain performance intact but causes degradation in the constraint OOD regime, identifying it as a key driver of cross-constraint generalization. Finally, replacing multi-task training with task-specific FODs marginally improves the BF rate but degrades the US rate and QoS satisfaction, indicating that the WFM consolidates pre-training tasks into a single shared backbone at negligible cost.

	\section{Conclusion}
	In this paper, we proposed a hierarchical WFM that decouples task-agnostic channel representation learning from task-conditioned decision-making for multi-task optimization. The proposed framework integrates an upstream FCE that learns universal channel representations with a downstream FOD that maps these representations to task-specific decisions via prompt-conditioned cross-attention. A hybrid supervised-to-unsupervised training scheme further shifts training from heuristic imitation toward direct performance maximization while a lightweight modular adaptation design extends the pre-trained backbone to channel estimation and beam selection with minimal parameter overhead. Simulation results demonstrate that the proposed framework learns high-fidelity channel representations, attains competitive multi-task optimization performance, and substantially reduces inference latency relative to numerical baselines. In addition, it generalizes robustly across unseen propagation environments, varying constraint parameters, and heterogeneous system configurations.
	Although this work focuses on TDD systems, the proposed framework is not inherently restricted to TDD operation. Extending it to FDD systems, where compact representations learned by the FCE may facilitate CSI feedback and uplink-to-downlink extrapolation, is a promising direction.


\end{document}